\documentclass[onecolumn,useAMS,usenatbib]{mn2e}

\usepackage{color}
\usepackage{amsmath}
\usepackage{graphicx}
\usepackage{amssymb}
\usepackage{psfrag}

\newcommand{\beq}{\begin{equation}}
\newcommand{\eeq}{\end{equation}}
\newcommand{\beqy}{\begin{eqnarray}}
\newcommand{\eeqy}{\end{eqnarray}}
\newcommand{\bay}{\begin{array}}
\newcommand{\eay}{\end{array}}
\renewcommand{\div}{\nabla\cdot}
\newcommand{\curl}{\nabla\times}
\newcommand{\brac}[1]{\left( {#1} \right)}

\newcommand{\pd}[2]{\frac{\partial{#1}}{\partial{#2}}}
\newcommand{\td}[2]{\frac{d{#1}}{d{#2}}}

\newcommand{\bB}{{\bf B}}
\newcommand{\bv}{{\bf v}}
\newcommand{\br}{{\bf r}}

\newcommand{\boldf}{{\bf f}}
\renewcommand{\bbeta}{\boldsymbol{\beta}}
\renewcommand{\bxi}{\boldsymbol{\xi}}
\newcommand{\bOm}{\boldsymbol{\Omega}}

\newcommand{\dP}{\delta P}
\newcommand{\drh}{\delta\rho}

\newcommand{\dB}{\delta\bB}

\newcommand{\Bav}{<\!\! B \!\!>}

\newcommand{\clD}{\mathcal{D}}

\newcommand{\dvmu}{\frac{1}{4\pi}} 

\title[Oscillations of rotating magnetised NSs with toroidal
fields]{Oscillations of rotating magnetised neutron stars with purely toroidal
  magnetic fields}
\author[S. K. Lander, D. I. Jones and A. Passamonti]
       {S. K. Lander\thanks{skl@soton.ac.uk},
        D. I. Jones\thanks{d.i.jones@soton.ac.uk} and
        A. Passamonti\\
University of Southampton, Southampton, U. K.}
\begin{document}


\pagerange{\pageref{firstpage}--\pageref{lastpage}} \pubyear{0000}

\maketitle

\label{firstpage}

\begin{abstract}
We investigate the oscillation spectrum of rotating Newtonian neutron stars
endowed with purely toroidal magnetic fields, using a time evolution
code to evolve linear perturbations in the Cowling approximation. The
background star is generated by numerically solving the MHD
equilibrium equations and may be nonspherical by virtue of both
rotation and magnetic effects; 
hence our perturbations and background are fully consistent. Whilst
the background field is purely toroidal, the perturbed field is mixed
poloidal-toroidal. From Fourier analysis of the perturbations we are
able to identify a number of magnetically-restored Alfv\'en (or $a$-)
modes. We show that in a rotating star pure inertial and $a$-modes are
replaced by hybrid magneto-inertial modes, which reduce 
to $a$-modes in the nonrotating limit and inertial modes in the
nonmagnetic limit. We show that the $r$-mode instability is suppressed
by magnetic fields in sufficiently slowly rotating stars. In addition, we 
determine magnetic frequency shifts in the $f$-mode. We discuss the
astrophysical relevance of our results, in particular for magnetar
oscillations.
\end{abstract}

\begin{keywords}
MHD --- stars: magnetic fields --- stars: neutron --- stars:
oscillations --- stars: rotation
\end{keywords}

\section{Introduction}

The identification of Soft Gamma Repeaters (SGRs) and Anomalous X-ray
Pulsars (AXPs) as magnetars has reignited interest in the role played 
by magnetic fields in stellar dynamics. A class of neutron star (NS) with
surface fields of $\sim 10^{15}$ gauss, magnetars are the most highly
magnetised stars yet observed. The SGRs are of special interest since
in addition to their regular gamma-ray flares, they are also
susceptible to occasional 
giant flares over three orders of magnitude more energetic than the
regular flares. In the tails of these
flares, high-frequency quasi-periodic oscillations (QPOs) have been
observed \citep{israel,watts_stroh} --- these may be the first direct
detections of neutron star 
oscillations. These QPOs should provide a valuable probe of the
physics of the NS interior, but this requires an understanding of
stellar oscillations in the presence of a strong magnetic field.

Neutron stars are not the only stars where magnetic effects
may be important. The influence of a star's magnetic field on its
oscillation spectrum can be gauged from the ratio of its magnetic
energy to the gravitational binding energy, $M/|W|$; this suggests
three classes of star where one should take account of the star's
magnetic field: in addition to NSs, there are also the
rapidly-oscillating type-A peculiar (roAp) stars and magnetic white
dwarfs (MWDs).

The earliest studies of magnetic star
oscillations were driven by the discovery of $\sim 10^4$ gauss fields
--- relatively strong for a main-sequence star --- in some Ap stars
\citep{chandlimb,ledoux}. Later, some of these stars were found to be
oscillating at high frequency --- the roAp stars --- motivating a
number of studies of magnetic effects on high-frequency $p$-modes
\citep{unno,dziem,rinc_rieu}.  In addition to roAp stars,
some white dwarfs have strong ($\sim 10^9$ gauss) magnetic fields;
these have shown no evidence of pulsation, perhaps due to magnetic
suppression of the $g$-modes observed in weaker-field white
dwarfs\citep{wickrama}. Finally, the internal dynamics of neutron
stars will be 
considerably affected by rotation as well as their strong magnetic
fields, leading to interest in magnetic $r$-modes \citep{morsink}.

Many publications to date have reported on analytic studies of
magnetic stellar oscillations, necessitating considerable
simplications to the problem: typically the model used is an
incompressible star with a force-free background magnetic field. Some
modern work on the problem has been inspired by the observation of
magnetar QPOs, and this has tended to be numerical
\citep{sotani08,sotani09,cerda}, with the advantages that more
sophisticated physics can be modelled (for example, compressible and
relativistic stars).

This paper is an investigation of the mode spectrum of magnetic stars
through time evolution of the perturbed MHD equations. Here we
consider only purely toroidal background fields, but in future we
intend to investigate purely poloidal and mixed field stars too. We begin by
perturbing the full system of equations to yield a set of equations
for the background and another set for the perturbations. We improve
on previous work on magnetic oscillations by solving the
background equations in a self-consistent manner, so that the star is
in exact equilibrium (up to some small numerical error), rather than
simply using spherically symmetric density profiles and analytic
results for the magnetic field
configurations. We next discuss details of our time-evolution code and
test its accuracy. After this, we present results for
oscillations of stars with magnetic fields and rotation. Finally, we
interpret these in the light of observed magnetic star oscillations
and compare them with previous work on this subject.

\section{Background and perturbation equations}

We model a neutron star as a self-gravitating, rotating, magnetised
polytropic fluid with perfect conductivity. The system is then
governed by the equations of ideal magnetohydrodynamics (MHD):
\beq
\rho\brac{\pd{\bv}{t} + (\bv\cdot\nabla)\bv + 2\bOm\times\bv}
  = -\nabla P - \rho\nabla\Phi - \rho\bOm\times(\bOm\times\br)
    + \dvmu(\curl\bB)\times\bB,
\eeq
\beq
\nabla^2\Phi = 4\pi G\rho,
\eeq
\beq \label{drhodt}
\pd{\rho}{t} = -\nabla\cdot(\rho\bv),
\eeq
\beq \label{dBdt}
\pd{\bB}{t} = \curl (\bv\times\bB),
\eeq
\beq
P=k\rho^\gamma,
\eeq
together with the solenoidal constraint $\div\bB=0$ on the magnetic
field. Here $\bv$ denotes the part of the fluid's velocity field which 
is not rigid rotation $\bOm$; all other symbols have their usual
meanings. Throughout this paper we work with $\gamma=2$ polytropes
exclusively, as a simple approximation to a neutron star equation of
state. We consider linear Eulerian perturbations of this system by 
making the standard ansatz that each physical quantity has a
zeroth-order background piece and a first-order perturbed piece;
e.g. the density is written as $\rho=\rho_0+\drh$.

We assume that our background star is stationary and rigidly rotating,
so that $\bOm$ is zeroth-order and $\bv$ first-order. Equations
\eqref{drhodt} and \eqref{dBdt} become trivial and we are left with
\beq \label{backeuler}
0 = -\nabla P_0 - \rho_0\nabla\Phi_0
    - \rho_0\bOm\times(\bOm\times\br) + \dvmu(\curl\bB_0)\times\bB_0,
\eeq
\beq
\nabla^2\Phi_0 = 4\pi G\rho_0,
\eeq
\beq
P_0=k\rho_0^\gamma.
\eeq
Making the additional assumption of axisymmetry one may show that this
system of equations splits into two cases: one where the magnetic
field is purely toroidal and a second mixed-field case (with
pure-poloidal fields as a limiting case). Details of the solution of
these equations are given in \citet{landerjones}; we use the code
described therein to generate the background configurations used
here. Here we merely note that our background configurations are fully
self-consistent, with rotation, magnetic fields and fluid effects in
equilibrium. In contrast to other work on magnetic oscillations, our
background star need not be spherical, but may be distorted by
rotational or magnetic effects, or a combination thereof.

For the perturbation equations, we work in the rotating frame of the
background and write our equations in terms of the perturbed density
$\drh$, the mass flux $\boldf=\rho_0\bv$ and a magnetic variable
$\bbeta=\rho_0\dB$. We additionally make the Cowling approximation ---
neglecting the perturbed gravitational force --- to avoid the
computational expense of solving the perturbed Poisson equation. Our
perturbations are then governed by seven equations:
\beq \label{euler_magmode}
\rho_0\pd{\boldf}{t}
  = -\gamma P_0\nabla\drh - 2\Omega\times\boldf
    + \brac{\frac{(2-\gamma)\gamma P_0}{\rho_0}\nabla\rho_0
                            - \dvmu(\curl\bB_0)\times\bB_0}\drh
    + \dvmu(\curl\bB_0)\times\bbeta + \dvmu(\curl\bbeta)\times\bB_0
    - \frac{1}{4\pi\rho_0}(\nabla\rho_0\times\bbeta)\times\bB_0
\eeq
\beq \label{conti_magmode}
\pd{\drh}{t}=-\div\boldf,
\eeq
\beq \label{induc_magmode}
\pd{\bbeta}{t} = \curl(\boldf\times\bB_0)
                 -\frac{\nabla\rho_0}{\rho_0}\times(\boldf\times\bB_0).
\eeq
If we rewrite these equations in terms of $\dP=\frac{\gamma
  P_0}{\rho_0}\drh$ and set the magnetic field to zero they reduce to
the perturbation equations of \citet{jones_evol}.

We next decompose each perturbed quantity in the azimuthal angle
$\phi$; for example $\drh$ becomes:
\beq
\drh(t,r,\theta,\phi)
   = \sum\limits_{m=0}^\infty
                               \drh_m^+(t,r,\theta) \cos m\phi
                              +\drh_m^-(t,r,\theta) \sin m\phi.
\eeq
This reduces our problem from a 3D to a 2D one, at the expense of
doubling the number of equations: we now have evolution equations in
fourteen variables:
$f_r^\pm,f_\theta^\pm,f_\phi^\pm,\drh^\pm,\beta_r^\pm,\beta_\theta^\pm,\beta_\phi^\pm$
for some value of $m$ to be specified at the start of the evolution.

\subsection{Boundary conditions}

Rotational and magnetic forces will serve to distort the star's
density distribution away from spherical symmetry and hence complicate
the treatment of perturbations at the stellar surface. To avoid these
complications we replace the radial coordinate $r$ with one fitted to
isopycnic surfaces, $x=x(r,\theta)$; even a nonspherical surface will
be defined by one value $x\equiv R$. With the
background density being a function of $x$ alone, we have
$\rho_0(x\!=\!R)=0$ and hence
\beq
\boldf(x\!=\!R)=\bbeta(x\!=\!R)={\bf 0}.
\eeq
Finally, the Lagrangian pressure perturbation $\Delta P$ is zero at
the surface by definition. Relating this to the Eulerian perturbation
we have
\beq
\dP + {\boldsymbol\xi}\cdot\nabla P_0 = 0 \textrm{\ \ at the surface.}
\eeq
Using \eqref{backeuler}, we see that $\nabla P_0$ may be written as two
terms proportional to $\rho_0$ and a term involving the magnetic
current $\curl\bB/4\pi$. Both density and current are zero at
the stellar surface 
and so $\nabla P_0$ must also vanish there. This yields our last
surface boundary condition:
\beq
\delta P(x\!=\!R) = 0.
\eeq

Our boundary conditions allow us to evolve the interior magnetic field
perturbations of our star, but not oscillations of the exterior. By
contrast, one would expect magnetic perturbations in a physical
neutron star to reach the surface and produce electromagnetic
radiation extending through the exterior. Whilst our treatment of the
surface does not account for this, we believe that it is the most that
can be done using the equations of perfect MHD: in an
infinitely-conducting polytropic star, a magnetic field that extends
to the surface has a corresponding Alfv\'en speed
$c_A\equiv\sqrt{B^2/4\pi\rho}$ which becomes superluminal as $\rho\to
0$, and infinite when $\rho=0$ (i.e. the stellar surface and
exterior). Dealing with the surface
and exterior thus requires extra physics: a stellar model more
sophisticated than a polytropic fluid with perfectly electrical
conductivity. One could employ
a low-density numerical atmosphere for the exterior, or assume that
the field is confined or matches to some simplified crust --- but
these are merely numerical conveniences rather than good models of
actual NS physics. In reality, perfect MHD ceases to be a good
approximation close to the surface of a NS, where resistive effects
become important and the full equations of electromagnetism should be
used. The stellar surface is not fluid but an elastic crust; and the
exterior will have a magnetosphere region rather than a dilute, uniform
`atmosphere'.

Needless to say, a credible model star which included all these
effects would give an oscillation spectrum closer to that of a real
neutron star than the one we study here. In lieu of such a model,
however, we treat oscillations over the 
fluid, highly-conductive interior of the star only. With magnetic
fields being strongest here and $\sim 99\%$ of the NS's mass
consisting of a fluid interior, we anticipate that dynamics in this
region would dominate the star's oscillation spectrum; and hence that
our treatment is a reasonable first attempt to understand oscillations
in real NSs. 

Next we look at the conditions at the centre of the star. Since we
deal with $m>0$ perturbations in this study, we should enforce a
zero-displacement condition:
\beq
\dP(x\!=\!0)=0\ ,\ \boldf(x\!=\!0)=\bbeta(x\!=\!0)={\bf 0}.
\eeq

In general, a parity analysis of the perturbation equations of a fluid
star shows that the variables may be classed according to their
equatorial symmetry --- either odd (the perturbation is zero
along the equator) or even (its theta-derivative is zero). This
division of the perturbations allows us to reduce our numerical domain
to just one 2D quadrant and enforce the perturbation symmetry at the
equator as another set of boundary conditions.

Analysing the perturbation equations for the (unmagnetised) rotating fluid
problem, one finds that the perturbation variables may be divided
into the two symmetry classes $\{f_r^\pm,f_\phi^\pm,\drh^\pm\}$ and
$\{f_\theta^\pm\}$. In the case
of a background star with a pure poloidal field these classes are
augmented by magnetic variables,
viz. $\{f_r^\pm,f_\phi^\pm,\drh^\pm,\beta_\theta^\pm\}$,
$\{f_\theta^\pm,\beta_r^\pm,\beta_\phi^\pm\}$.
Note that although the background field is pure-poloidal, the
perturbed field will still be mixed poloidal-toroidal. For a pure-toroidal
background the magnetic perturbations are again mixed, but they fall
into \emph{different} symmetry classes from perturbations of a
pure-poloidal star:
$\{f_r^\pm,f_\phi^\pm,\drh^\pm,\beta_r^\pm,\beta_\phi^\pm\}$ and 
$\{f_\theta^\pm,\beta_\theta^\pm\}$.
It follows that whilst we may separately treat perturbations on either
a pure-poloidal or pure-toroidal background, the perturbations of a
mixed-field background will have no definite equatorial
symmetry. Investigating this latter group of perturbations requires an
extended numerical domain consisting of an upper and lower
quadrant. For this paper we concentrate only on oscillations of stars
with purely toroidal background fields, postponing the pure-poloidal
and mixed-field cases to future work.

\section{Numerics}

As described above, our numerical domain is one 2D quadrant of a
circle, with $x\in [0,1]$ and $\theta\in [0,\pi/2]$; by symmetry and
through a $\phi$-decomposition this domain is sufficient to
investigate behaviour over the whole 3D, potentially nonspherical,
star. Upon decomposing in $\phi$, equations \eqref{euler_magmode},
\eqref{conti_magmode} and \eqref{induc_magmode} become a system of
fourteen perturbation equations, which we evolve using a MacCormack
predictor-corrector algorithm.

Our code is an extended version of the
one described in \citet{passa_strat}, where the authors demonstrated
its accuracy and long-term stability for barotropic and stratified
stars, in the nonmagnetic case. As in their work, we employ a
fourth-order Kreiss-Oliger dissipation; this is an extra term added to
the equations to damp spurious higher-order oscillations, i.e. those
generated by the code's finite differencing. The magnitude of this
term is resolution-dependent, so that it vanishes in the
infinite-resolution continuum limit.

In addition to this dissipation, two further tricks are required to
ensure stability and accuracy of magnetic evolutions. To stabilise the
numerical evolution of the magnetic field, we first note that if the
electrical resistivity $\eta$ is non-zero, the induction equation
gains an extra term:
\beq \label{resist_induc}
\pd{\bB}{t} = \curl(\bv\times\bB) - \eta\curl(\curl\bB).
\eeq
By including this second term (at a small magnitude) we are able to
suppress instabilities which arise from evolving the magnetic
field. As for the Kreiss-Oliger dissipation, this artificial
resistivity is added in a resolution-dependent
manner, so it reduces to the correct (i.e. physical) 
continuum limit. We find that a very small value of $\eta$ is
sufficient to improve long-term stability, but has negligible physical
effect on our evolutions, since it acts over a far longer timescale
than any others in our problem.

\subsection{Divergence cleaning}

Finally, for the long-term accuracy of the code we need to ensure that
the perturbed magnetic field remains solenoidal. This is guaranteed in
the continuum limit, since the divergence of the induction equation is
\beq
\pd{(\div\bB)}{t} = \div\curl(\bv\times\bB) \equiv 0,
\eeq
but in practice numerical error will be introduced from the finite
grid resolution. It is important to `clean' the field of this class of
numerical error, since it has been shown that a numerically-generated
monopolar field gives rise to a spurious extra force \citep{brackbarnes}.

There are various approaches to divergence cleaning for numerical
schemes. A review of these may be found in \citet{dedner}, where in
addition a new constrained formulation of MHD is proposed, where the
condition $\div\bB=0$ is coupled to the induction equation through an
auxiliary function; we repeat their argument below.

In the continuum limit the induction equation states that the vector
$\partial_t\bB$ has a divergence-free part only, whereas a general
vector can be decomposed into curl-free and divergence-free parts. Our
discretised induction equation will no longer preserve this
divergence-free aspect exactly and accordingly we add a curl-free
term $-\nabla\psi$ to the RHS, with $\psi$ being some unknown
function. We then couple our augmented induction equation to a
relation for $\psi$:
\beqy \label{divclean_induc}
\partial_t\bB &=& \curl(\bv\times\bB) - \nabla\psi \\
\clD(\psi) &=& -\div\bB
\eeqy
where $\clD$ is some linear differential operator. The Euler
equation and the equation of mass conservation are unaffected. We now
take the divergence of the first relation and the time derivative of
the second:
\beqy
\partial_t(\div\bB)  &=& -\nabla^2\psi \\
\partial_t\clD(\psi) &=& -\partial_t(\div\bB)
\eeqy
which we combine to see that
\beq \label{clDpsi+psi}
\partial_t\clD(\psi) = \nabla^2 \psi.
\eeq

The choice of $\mathcal{D}$ determines the way in which divergence
errors are removed. The three basic types of cleaning are elliptic,
parabolic and hyperbolic --- so named because they entail solving a
heat equation, wave equation or Poisson equation,
respectively. \citet{dedner} pioneer a mixed hyperbolic-parabolic
approach, which they find to be superior to the simpler
divergence-cleaning methods since it allows for errors to be
propagated out of the star (hyperbolic cleaning) whilst simultaneously
being damped (parabolic cleaning). The third method, elliptic
cleaning, has the serious disadvantage that it requires the repeated
solution of the (computationally expensive) Poisson equation; the
mixed-hyperbolic scheme only adds the modest expense of having to
evolve one more quantity --- the function $\psi$.

Hyperbolic-parabolic divergence cleaning involves defining
$\mathcal{D}$ by
\beq
\mathcal{D}(\psi) = \frac{1}{c_h^2}\partial_t\psi + \frac{1}{c_p^2}\psi,
\eeq
which leads to a telegraph (damped-wave) equation for $\psi$:
\beq
\partial_{tt}\psi = -\frac{c_h^2}{c_p^2}\partial_t\psi
                     + c_h^2 \nabla^2 \psi.
\eeq
Within the code, we implement this divergence-cleaning method through
the evolution equation
\beq
\partial_t\psi = -\frac{c_h^2}{c_p^2}\psi - c_h^2\div\bB
\eeq
together with our modified induction equation
\eqref{divclean_induc}. Following \citet{pricemona} we take $c_h$, the
divergence-wave propagation speed, to be related to the sound $c_s$
and Alfv\'en $c_A$ speeds through the relation:
\beq
c_h = \sqrt{c_s^2+c_A^2}.
\eeq
The other coefficient is physically the inverse of the decay timescale
$\tau$:
\beq
\frac{c_h^2}{c_p^2} = \frac{1}{\tau}
\eeq
which \citet{pricemona} argue is not universal, but rather should be
modified to suit some lengthscale $\lambda$ specific to the problem, i.e.
\beq
\frac{c_h^2}{c_p^2} = \frac{1}{\tau} = \frac{\alpha c_h}{\lambda},
\eeq
where $\alpha$ is a dimensionless parameter. Using this result, we take
$\lambda$ to be the radial grid spacing $\Delta r$ in our
code. Finally then, our evolution equation for the function $\psi$ is
\beq
\partial_t\psi = -\frac{\alpha\sqrt{c_s^2+c_A^2}}{\Delta r}\psi
                  - (c_s^2+c_A^2)\div\bB.
\eeq

To close the system we need to give appropriate boundary conditions
and initial data. For the latter we simply set $\psi(t=0)=0$ --- this
is reasonable because the initial data is divergence-free and so the
variable $\psi$, associated with the monopole part of the magnetic
field, should be zero initially.

For the boundary condition at the surface, we choose the Sommerfeld
outgoing wave condition on $\psi$:
\beq
\partial_t\psi = -\sqrt{c_s^2+c_A^2} \brac{\partial_r\psi+\psi}.
\eeq
This result is for a spherical surface, but it still gives 
satisfactory cleaning in the case where the background star is
spheroidal.

\subsection{Testing the code}

Since we already have confidence in the performance of the code in the
nonmagnetic limit (see \citet{passa_strat} for details), we now test
its accuracy and convergence properties with the inclusion of magnetic
effects. To this end, we wish to monitor the divergence of the
magnetic field and the total energy of the system (which should be
conserved in the continuum limit). Since the background is stationary
its total energy is automatically conserved; in addition the
background field was constructed in a divergence-free manner (see
\citet{landerjones}). Therefore it suffices to check conservation of
the perturbed energy and the value of $\div\dB$.

To get a measure of the divergence of $\dB$ over the \emph{whole}
star, rather than at individual points, we would like to evaluate
$\div\dB$ volume-integrated over the star. However, this quantity is
first-order and so integrates to zero, so instead we define a
`monopole energy'
\beq
\mathfrak{D}\equiv\frac{R^2}{8\pi}\int \left(\div\dB\right)^2 dV
\eeq
where $R$ is the stellar radius; this energy should be insignificant
in comparison with the perturbed magnetic energy $\delta M$. For each
evolution used to generate results for this paper, the divergence of
$\dB$ was monitored through 
the dimensionless quantity $\mathfrak{D}/\delta M$. Typically this
quantity was found to be of the order $\sim 0.01$, rising to $\sim
0.1$ in the case of very strong fields and rapid rotation rates.

Next we use conservation of (perturbed) energy to test the order of
convergence of our code, using the fact that in the limit of infinite
resolution energy should be exactly conserved. The total energy
$\delta\mathfrak{E}$ of the perturbations is given by
\beq
\delta\mathfrak{E} = \delta T+\delta W+\delta U+\delta M,
\eeq
i.e. the sum of the perturbed kinetic $T$, gravitational $W$, internal
$U$ and magnetic $M$ energies. Since we are making the Cowling
approximation, $\delta W=0$ and we are left with
\beq
\delta\mathfrak{E} = \delta T+\delta U+\delta M
                   = \int \brac{ \frac{1}{2}\rho_0 |\bv|^2
                                 + \frac{\gamma p_0}{2\rho_0^2}\drh^2
                                 + \frac{1}{8\pi} |\dB|^2 }\ dV.
\eeq
This is in agreement with equation (C5) of \citet{friedmanschutz} in
the case of adiabatic perturbations within the Cowling approximation,
but with an additional magnetic energy term. Note that all these terms
are second-order quantities; the first-order energy vanishes for a
background in equilibrium, and in our case each piece of it (for
example the magnetic energy term proportional to $\bB_0\cdot\dB$) is
automatically zero by virtue of our $\phi$-decomposition.

\begin{figure}
\begin{center}
\begin{minipage}[c]{\linewidth}
\psfrag{etot}{$\displaystyle\frac{\delta\mathfrak{E}(t)}{\delta\mathfrak{E}(0)}$}
\psfrag{t}{$t$}
\psfrag{conv}{$\mathcal{O}_{conv}$}
\psfrag{cont}{continuum}
\psfrag{med}{$32\times 30$}
\psfrag{high}{$64\times 60$}
\includegraphics[width=\linewidth]{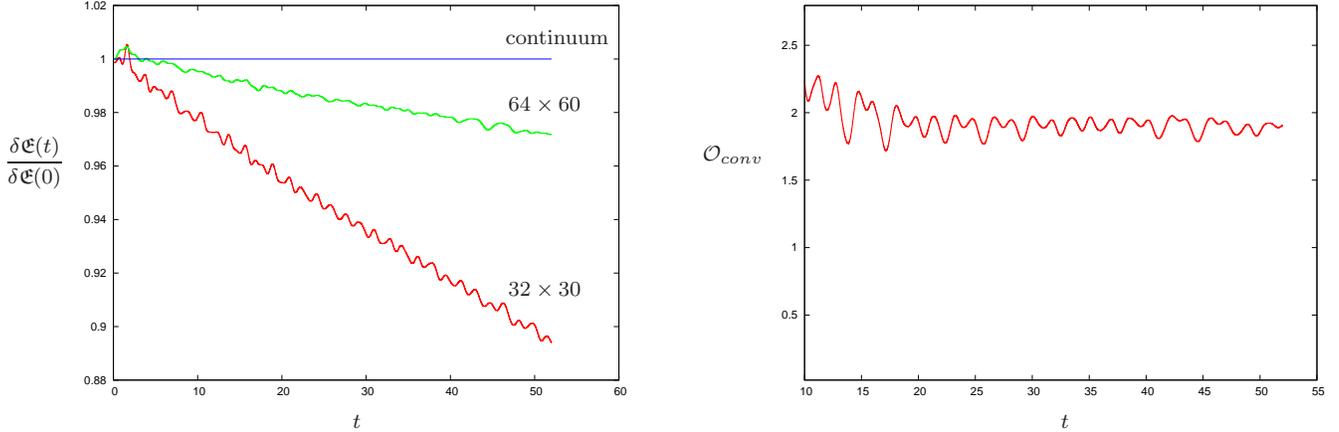}
\end{minipage}
\caption{\label{conv}
         We determine the order of convergence of our code by
         evaluating the total perturbed energy $\delta\mathfrak{E}$
         over time; in the exact, continuum limit this quantity will
         not deviate from its initial value. The left-hand plot shows
         the deviation of $\delta\mathfrak{E}$ from its initial value
         for $(r,\theta)$ grids of $32\times 30$ and $64\times
         60$ points. From these we confirm that the order of convergence
         $\mathcal{O}_{conv}$ of the code is equal to 2, as intended (see
         right-hand plot). $\mathcal{O}_{conv}$ is only plotted for
         $t\geq 10$, since at early times the numerical values of
         $\delta\mathfrak{E}$ cross the continuum value, causing
         $\mathcal{O}_{conv}$ to oscillate rapidly. The background
         configuration for 
         these tests was a star with rotation rate
         $\Omega/\sqrt{G\rho}=0.238$ and with an average magnetic
         field strength $\Bav=2.87\times 10^{16}$ G, evolved for 30
         $f$-mode oscillations.}
\end{center}
\end{figure}

To evaluate the convergence ratio, we monitor the evolution of the
high-resolution energy $\delta\mathfrak{E}_{64\times 60}(t)$ and the
medium-resolution energy $\delta\mathfrak{E}_{32\times 30}(t)$,
comparing these with the initial value of the energy
$\delta\mathfrak{E}(0)$. In the continuum limit $\delta\mathfrak{E}$
will have no time-dependence and will be equal to its initial value
for all time. Hence we are able to use this exact result to define a
convergence ratio
\beq
\mathcal{O}_{conv}
 = \frac{1}{\log 2}\log\brac{
    \frac{\delta\mathfrak{E}_{32\times 30}(t)-\delta\mathfrak{E}(0)}
         {\delta\mathfrak{E}_{64\times 60}(t)-\delta\mathfrak{E}(0)} }.
\eeq
In figure \ref{conv} we evaluate $\mathcal{O}_{conv}$ over time,
confirming that the code is second-order convergent.

\subsection{Nondimensionalising}

Throughout the code we employ variables which have been made
dimensionless through division by a suitable combination of powers of
gravitational constant $G$, central density $\rho_c$ and equatorial
radius $r_{eq}$. For example, a dimensionless mode frequency
$\hat\sigma$ is related to the physical one $\sigma$ (with units of
rad s${}^{-1}$) through the
relation $\hat\sigma=\sigma/\sqrt{G\rho_c}$; the conversion is the
same for rotational frequency $\Omega$. Since dimensionless
frequencies of this form are common in oscillation mode literature we
use these throughout this paper. Dimensionless magnetic field
strengths, however, are less likely to be familiar and so we quote 
these in terms of gauss.

When we use dimensional quantities they are for a neutron star with
canonical parameters: an equatorial radius of 10km (in the
non-rotating, unmagnetised case) and a mass of $1.4M_\odot$ (where
$M_\odot$ is solar 
mass). The relationship between dimensionless frequencies
$\hat\sigma$ (equivalently $\hat\Omega$) and their physical
counterparts is only weakly dependent on $\Omega$ and $\bB$ --- and hence
is roughly linear, with
\beq \label{redim}
\sigma\textrm{[Hz]} \approx 1890\hat\sigma.
\eeq
Finally, we note that in our dimensionless units, the Keplerian
(break-up) velocity $\Omega_K\approx 0.72$. When we plot sequences of
modes in rotating stars, we typically track the modes up to
$\Omega/\Omega_K\approx 0.95$; that is, rotation rates $95\%$ of the
break-up velocity.

\section{Results}

\subsection{Classes of oscillation mode}

Using spherical polar coordinates, a general perturbation may be
decomposed with respect to the basis 
($Y_{lm}{\bf e}_r,r\nabla Y_{lm},{\bf e}_r\times\nabla Y_{lm}$), where
$Y_{lm}=Y_{lm}(\theta,\phi)$ are the usual spherical harmonics. The
first two of these terms transform by multiplication by $(-1)^l$ under
parity inversion ${\bf r\mapsto -r}$, with the latter one transforming as
$(-1)^{l+1}$. This enables us to classify modes based on their
parity: those whose sign is given by $(-1)^l$ under
parity interchange as termed \emph{polar} modes, whilst those
transforming as $(-1)^{l+1}$ are called \emph{axial} modes. Hybrid
modes, consisting of a sum of axial and polar pieces, are termed
axial-led or polar-led based on 
whether the lowest-$l$ (i.e. $l\!=\!m$) term of the mode is axial or
polar, respectively.

For fixed $m$, \citet{lock_fried} found that inertial modes are not
characterised by a single $l$, but have an angular dependence
consisting of
a sum of spherical harmonics $Y_{lm}(\theta,\phi)$. However, in all
cases they found there was some threshold 
value $l_0$, such that the amplitude of $Y_{lm}$ contributions for $l>l_0$
was found to drop off rapidly. Following their work, we label
modes using the notation ${}^{l_0}_ma_k$, where $k$ distinguishes
between different modes with the same $l_0$.

A non-rotating, unmagnetised fluid star has only one class of
oscillation modes if the perturbations are assumed to have the same
equation of state as the background star. These are the $p$-modes,
whose restoring force is pressure fluctuations; the lowest-order
$p$-mode (i.e. the one with a nodeless eigenfunction) in each series
is termed the
fundamental mode, or $f$-mode. The non-axisymmetric $p$-modes are
degenerate in the absence of rotation and magnetic fields; each
$p$-mode has the same frequency for fixed $m$. These modes are polar
in nature.

With a rotating background star, a Coriolis force term enters the
equations governing the perturbations, which removes the
$m$-degeneracy in the $p$-modes. The
Coriolis term is the restoring force for a new class of modes: the
inertial modes, which we term $i$-modes. In general $i$-modes are mixed axial
and polar even in the slow-rotation limit, but one class of them are
purely axial in this limit: the $r$-modes. With the barotropic
equation of state we employ here, the only $r$-modes which exist are
those with $l\!=\!m$.

Finally, magnetic fields also induce a class of oscillation mode,
restored by the Lorentz force. We term them the Alfv\'en modes, or
$a$-modes. In addition to generating a new class of modes, the Lorentz
force can lift degeneracies of nonradial
oscillations, causing a splitting in mode frequencies
\citep{cox_book}. The addition of the Lorentz force term in the Euler
equation for the perturbations should produce shifts in the
frequencies of the $p,r$ and $i$ modes from their unmagnetised values.

\subsection{Initial data}

Having performed a $\phi$-decomposition of our perturbation variables,
we fix a value for the azimuthal index $m$ for each
evolution. However, since we have no restrictions on
$\theta$-dependence, oscillations for a variety of $l\geq m$ will
typically be excited for arbitrary initial data --- hence we are able to
study several modes at once. The nature of these modes is then
determined from analysis of their eigenfunctions, as well as by
tracking them back to a regime where we already know their properties.

Whilst any initial data will excite a variety of modes, we choose
different starting perturbations depending on whether we wish to
investigate axial/axial-led or polar/polar-led modes. For polar modes
we provide a density perturbation whose angular dependence is given
by an ordinary spherical harmonic; for axial modes we use a `magnetic'
spherical harmonic-dependent velocity perturbation. In both cases the
radial dependence is a Gaussian profile. More details about these
choices may be found in \citet{jones_evol}.

\subsection{Mode spectrum of a non-rotating magnetised star}

In this section we present results for nonrotating stars, since the mode
spectrum is simpler, leaving rotating stars to the next section. We
begin by investigating the new class of modes present with the 
addition of a magnetic field, the $a$-modes. Results are presented for
both polar and axial $a$-modes.

\begin{figure}
\begin{center}
\begin{minipage}[c]{0.7\linewidth}
\psfrag{PSD}{PSD}
\psfrag{nondim_sig}{$\displaystyle\frac{\sigma}{2\pi\sqrt{G\rho_c}}$}
\psfrag{noise}{noise}
\psfrag{f-mode}{$f$-mode}
\psfrag{a-modes}{$a$-modes}
\psfrag{mag}{magnetised}
\psfrag{unmag}{unmagnetised}
\includegraphics[width=\linewidth]{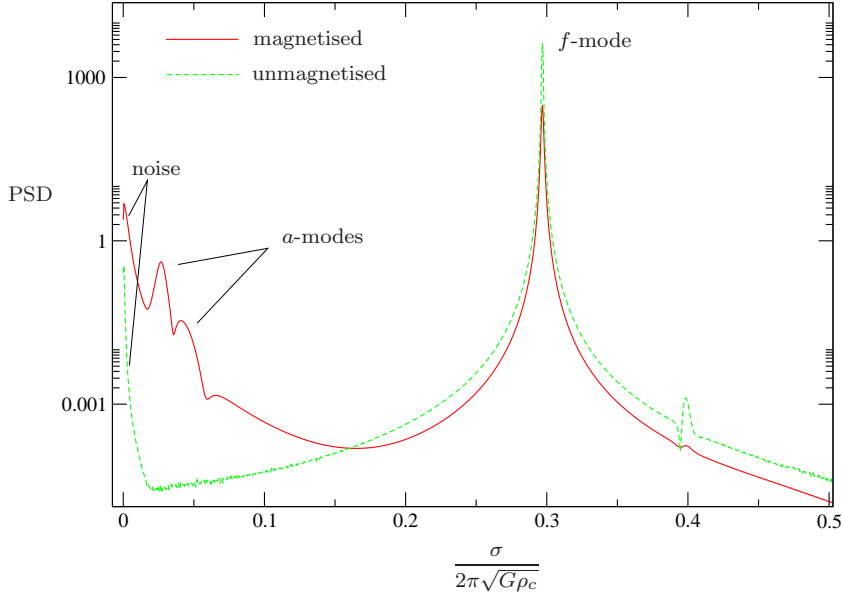}
\end{minipage}
\caption{\label{fft}
          Typical FFT results for a pair of nonrotating stars, one
          magnetised and the other unmagnetised. We plot mode
          frequency $\sigma$ (in a dimensionless 
          form) against PSD, the power spectral density. We see that the 
          $f$-mode frequencies are very close in both cases. With no
          Coriolis force there are no inertial modes, therefore any
          peaks at lower frequency than the $f$-mode must be either
          noise or Alfv\'en modes. We identify the lowest-frequency
          spike in the magnetic FFT as noise, since there is a
          corresponding unphysical peak in the nonmagnetic FFT. The
          following peaks in the magnetised-star FFT, however, have no
          analogue in the nonmagnetic FFT and so we identify these as
          Alfv\'en modes. The duration of the evolution was sufficient to
          resolve around 100 Alfv\'en oscillations.}
\end{center}
\end{figure}

Let us begin by considering where
in the frequency spectrum these modes could be expected. Now, any mode
frequency will be proportional to some characteristic wave speed. For
fluid modes like 
the $f$-mode, the frequency should be proportional to the sound speed
$c_s$; similarly the $a$-mode frequencies should be proportional to
the Alfv\'en speed $c_A$. Accordingly the ratio of frequencies should
scale as
\beq
\frac{\sigma_f}{\sigma_a} \sim \left< \frac{c_s}{c_A} \right>
\eeq
where the angle brackets represent a volume average. Now
\beq
\frac{c_s}{c_A} = \sqrt{\frac{\gamma P}{\rho}}
                              \bigg/ \sqrt{\frac{B^2}{4\pi\rho}}
\eeq
and so
\beq
\left< \frac{c_s}{c_A} \right> = \frac{2\sqrt{\pi\gamma <\!P\!>}}{\Bav}.
\eeq
We find from our background code that a nonrotating unmagnetised
$\gamma=2$ polytrope with a mass of $1.4M_\odot$ and radius $R=10$ km has
a volume-averaged pressure $<\!P\!>$ of $3.10\times 10^{34}$ dyn
cm${}^{-2}$. Using this value and $\Bav=10^{16}$ G to 
nondimensionalise, we find that
\beq
\frac{\sigma_f}{\sigma_a}
  \sim 90\times
     \brac{\frac{<\!P\!>}{3.10\times 10^{34} \textrm{ dyn cm}^{-2}}}^{1/2}
     \brac{\frac{\Bav}{10^{16} \textrm{ G}}}^{-1}.
\eeq
With the value of $<\!P\!>$ varying little with magnetic field
strength, we assume that it is a constant and that $\sigma_f/\sigma_a$
scales only with $\Bav$. It then follows that we should expect
$\sigma_a$ to be roughly  
$1/90$ of $\sigma_f$ for a $10^{16}$ G field, but $1/9$ of $\sigma_f$
for a $10^{17}$ G field. This part of the spectrum is dominated
by inertial modes in the case of unmagnetised rotating stars, but in
the absence of rotation 
we may be confident that any oscillations at lower frequency than the
$f$-mode are associated with the magnetic field --- see figure
\ref{fft}.

Now, with $\sigma_a\propto <\!c_A\!>$ and $c_A=B/\sqrt{4\pi\rho}$, it
follows that $\sigma_a\propto B$, provided that magnetic changes to
the density distribution are higher order (which should be true for
all but very high fields --- see \citet{landerjones}). To summarise,
$a$-modes should scale linearly with field strength and appear as
oscillations with lower frequency than the $f$-mode. With these
expectations, we now turn to numerical results from our time-evolution
code.

\begin{figure}
\begin{center}
\begin{minipage}[c]{\linewidth}
\psfrag{B}{$\Bav$ [gauss]}
\psfrag{sigma}{$\displaystyle\frac{\sigma}{\sqrt{G\rho_c}}$}
\psfrag{m=2}{$m\!=\!2$}
\psfrag{m=4}{$m\!=\!4$}
\psfrag{m=6}{$m\!=\!6$}
\psfrag{2a}{${}^2_2a$}
\psfrag{4a}{${}^4_4a$}
\psfrag{6a}{${}^6_6a$}
\psfrag{3a1}{${}^3_2a_1$}
\psfrag{3a2}{${}^3_2a_2$}
\psfrag{5a1}{${}^5_4a_1$}
\psfrag{5a2}{${}^5_4a_2$}
\psfrag{7a1}{${}^7_6a_1$}
\psfrag{7a2}{${}^7_6a_2$}
\includegraphics[width=\linewidth]{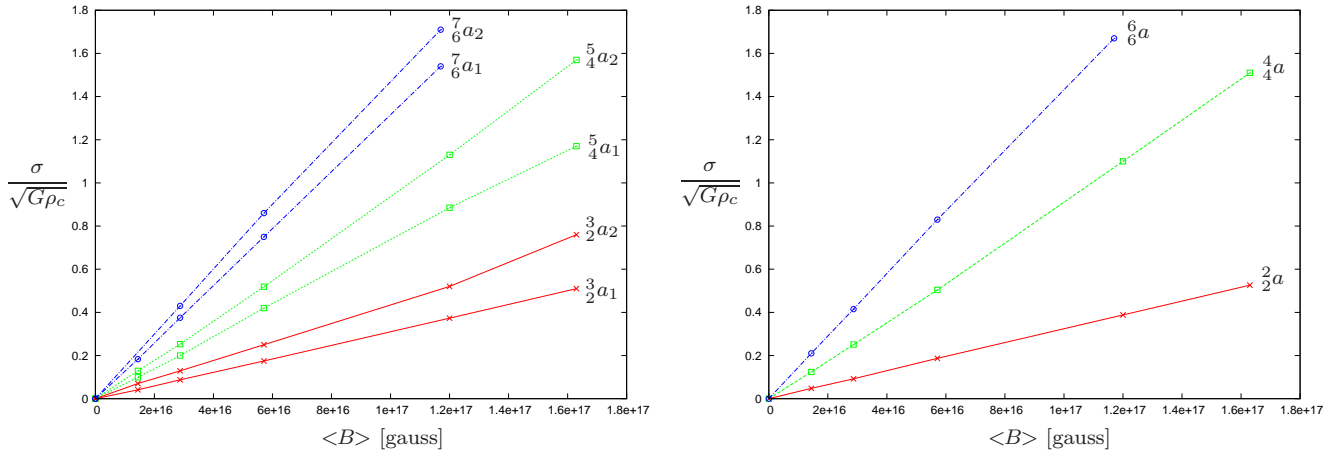}
\end{minipage}
\caption{\label{alf_mode}
         Polar $l_0=m+1$ Alfv\'en modes (left) and axial $l=m$ Alfv\'en
         modes (right), for
         $m\!=\!2,4,6$. Tracking the modes to high field strength, we
         see that each mode frequency scales linearly with magnetic
         field strength, as anticipated. These results are for a
         nonrotating star.}
\end{center}
\end{figure}

In figure \ref{alf_mode} we track a number of Alfv\'en mode
frequencies up to averaged-field strengths of order $10^{17}$
gauss. For axial initial data and fixed $m$ we find a single $l\!=\!m$
mode, whilst polar initial data excites two $l_0\!=\!m+1$ modes for a
given $m$. In all cases, we see that as expected there is a
near-linear relationship between $\sigma_a$ and $\Bav$. The
identification of the $a$-modes is based on analysis of their
eigenfunctions, using the numerical method of
\citet{ster_eigenfn}. The labelling used here anticipates the results
of the next section, where we track these modes for increasing
rotation rate.

At the start of this section we showed that the $a$-mode frequency
should vary linearly with $\Bav$, and this appears to be borne out by our
results. We now quantify this dependence and the deviation from
it. By looking at the weak-field results from our code (where the relationship
should be closest to linear), we determine the constants of
proportionality in the relationship
\beq \label{linear-sigB}
\frac{{}^{l_0}_m(\sigma_a)_k}{\sqrt{G\rho_c}} 
        = {}^{l_0}_m c_k \brac{\frac{\Bav}{10^{16}\textrm{ G}}},
\eeq
finding that ${}^2_2c=0.033,{}^3_2c_1=0.030,{}^3_2c_2=0.045,
{}^4_4c=0.086,{}^5_4c_1=0.069,{}^5_4c_2=0.090,
{}^6_6c=0.146,{}^7_6c_1=0.127,{}^7_6c_2=0.150$. For all our results
the linear relationship \eqref{linear-sigB}, with the
numerically-established constants ${}^{l_0}_m c_k$, agrees to within
8\% of the mode frequencies we have extracted from our evolutions ---
and in most cases the difference is less than 5\%.

\begin{figure}
\begin{center}
\begin{minipage}[c]{0.6\linewidth}
\psfrag{Bsq}{$\Bav^2$ [gauss${}^2$]}
\psfrag{shift}{\hspace{-0.5cm}shift in $\sigma_f$ (\%)}
\psfrag{m2}{$m=2$}
\psfrag{m4}{$m=4$}
\psfrag{m6}{$m=6$}
\includegraphics[width=\linewidth]{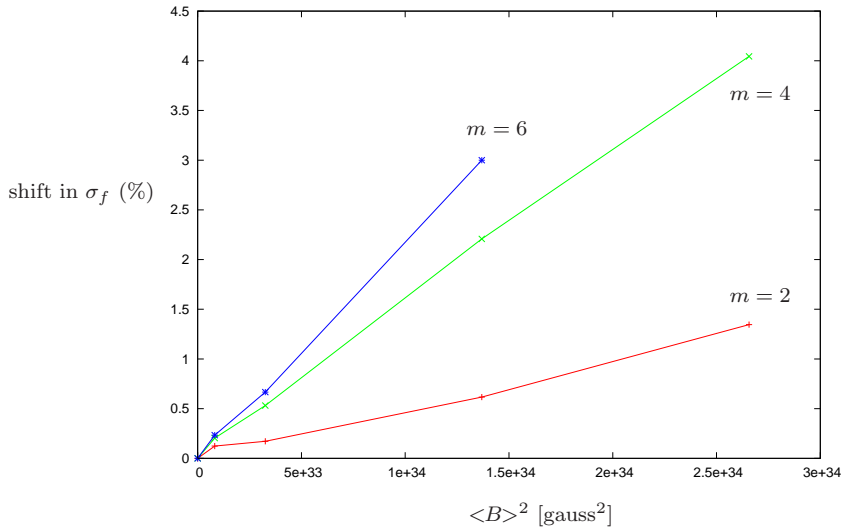}
\end{minipage}
\caption{\label{f_shift_norot}
         The shift in $f$-mode frequency due to magnetic effects (for
         nonrotating stars), for $m=2,4,6$. On the $y$-axis we plot
         percentage increase in 
         $\sigma_f$ from its unmagnetised value; we see that this
         shift appears to depend quadratically on $\Bav$. The apparent
         deviation from this dependence, visible in the weakest-field
         results, is attributable to numerical errors in these very
         small frequency shifts.}
\end{center}
\end{figure}

Finally in this section, we look at the shift in the frequency of the
fundamental mode upon the addition of a magnetic field to the
star. This mode is restored by perturbations in the fluid pressure $P$
in the unmagnetised case, so we anticipate that in the magnetic
problem the restoring force is perturbations of \emph{total} (fluid
plus magnetic)
pressure, $P+B^2/8\pi$. The magnetic shift in $\sigma_f$,
then, should be proportional to $B^2$ --- but since magnetic pressure
is very modest in magnitude compared with fluid pressure, we expect
the frequency shift to be small. For example, using our canonical
model star, the magnetic pressure is $\sim 1\%$ of the fluid pressure
at $10^{17}$ G. We confirm these expectations in
figure \ref{f_shift_norot}. In all cases $\sigma_f$ is increased by
the inclusion of magnetic effects, but the shifts are only around a
couple of percent even for $\Bav\sim 10^{17}$. The relative shift
appears to be more pronounced for higher-$m$ oscillations.

\subsection{Mode spectrum of a rotating magnetised star}

Armed with some idea of the mode spectrum of magnetised nonrotating
stars, we next consider rotating magnetised configurations. The
earliest studies of magnetic oscillations suggested that the
significance of the magnetic field on the oscillation spectrum should
be linked to the ratio $M/|W|$; when additionally including rotational
effects we would expect the relative significance of the two
effects to be related to $M/T$.

\begin{figure}
\begin{center}
\begin{minipage}[c]{0.5\linewidth}
\psfrag{sig}{$\displaystyle\frac{\sigma_f}{\sqrt{G\rho_c}}$}
\psfrag{omega}{$\displaystyle\frac{\Omega}{\sqrt{G\rho_c}}$}
\psfrag{mag}{magnetised}
\psfrag{unmag}{unmagnetised}
\includegraphics[width=\linewidth]{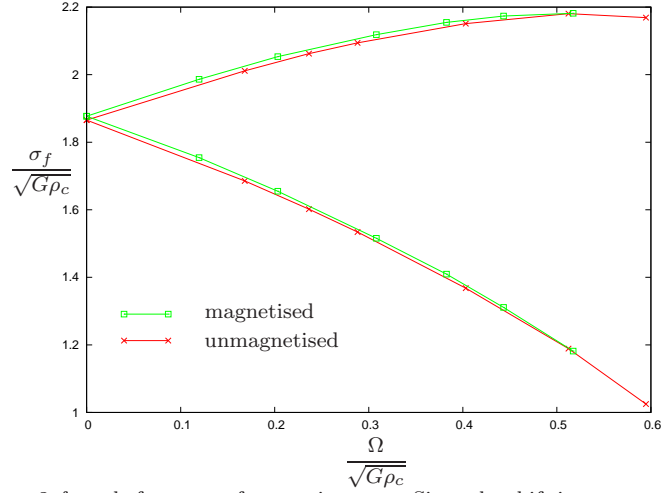}
\end{minipage}
\caption{\label{f_shift_rot}
         Magnetic shift of the $m=2$ $f$-mode frequency for rotating
         stars. Since the shift is very small we take a very highly
         magnetised background star, with $\Bav=1.17\times 10^{17}$ G
         for comparison with the nonmagnetic sequence of results. We
         find that as the rotation rate $\Omega$
         increases, magnetic effects become less significant.}
\end{center}
\end{figure}

We first consider magnetic shifts in the $f$-mode frequency for
rotating stars. Rotation splits the $f$-mode into co- and
counter-rotating modes; we expect the frequencies of both
branches of the mode to shift with the addition of magnetic fields. At
low rotation, the magnetic shift for
each piece of the $f$-mode is comparable with the shift in the
nonrotating case, but at higher rotation rates the shift becomes less
significant --- see figure \ref{f_shift_rot}. This bears out our
expectation that the magnetic shift should scale with $M/T$.

We next turn to $a$-modes and $r$-modes of rotating magnetised
stars. Based on our experience so far, we have expectations on how
each mode should behave. We anticipate a rotational splitting of the
$a$-modes into co- and counter-rotating pieces (as seen for the
$f$-mode); in addition we expect to see some magnetic shift, scaling
with $M/T$, in the $r$-mode.

\begin{figure}
\begin{center}
\begin{minipage}[c]{0.49\linewidth}
\psfrag{sigma}{$\displaystyle\frac{\sigma}{\sqrt{G\rho_c}}$}
\psfrag{omega}{$\displaystyle\frac{\Omega}{\sqrt{G\rho_c}}$}
\psfrag{rmode}{${}^2_2r$-mode}
\psfrag{amode}{${}^2_2a$-mode}
\includegraphics[width=\linewidth]{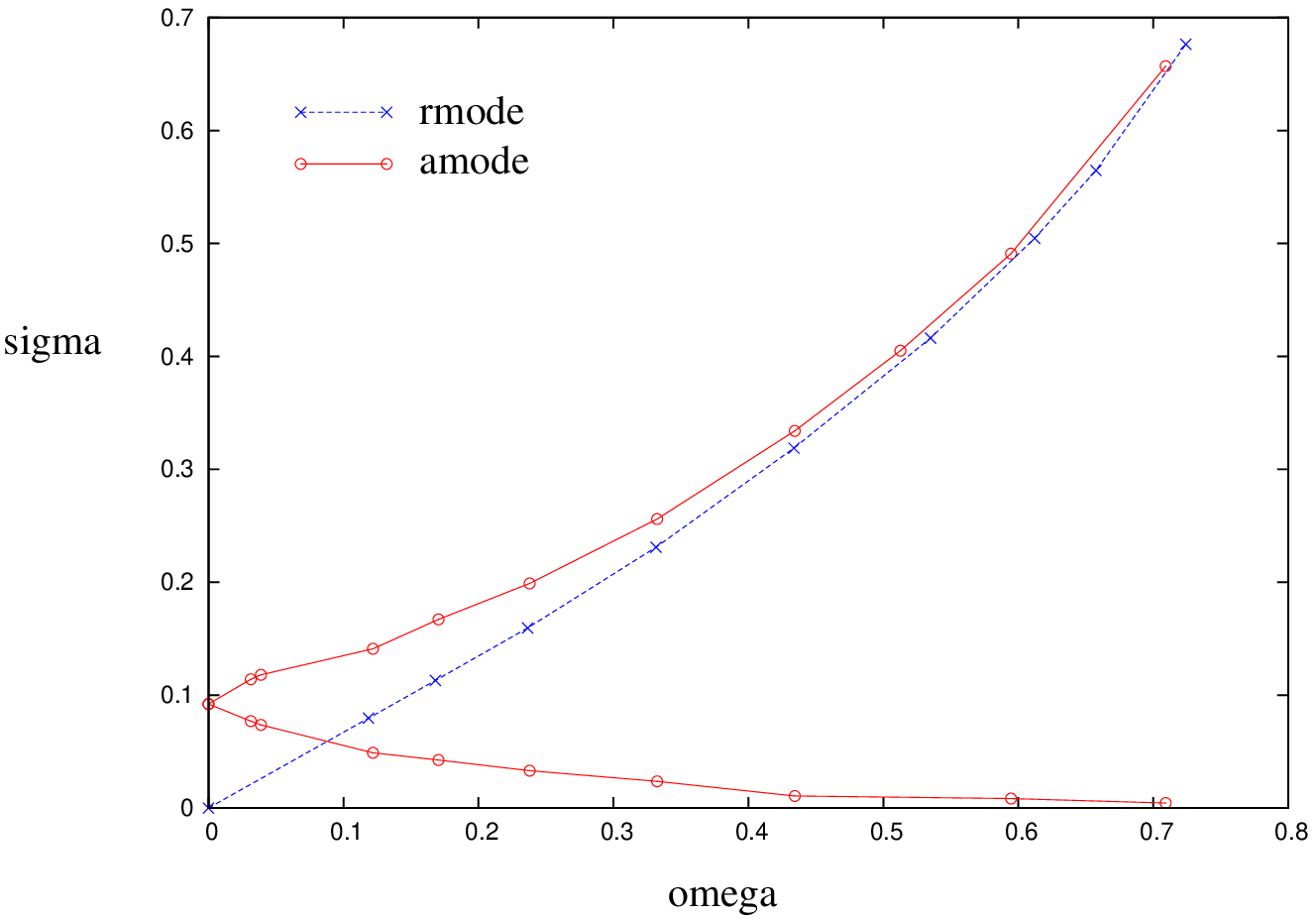}
\end{minipage}
\hfill
\begin{minipage}[c]{0.49\linewidth}
\psfrag{sigma}{$\displaystyle\frac{\sigma}{\sqrt{G\rho_c}}$}
\psfrag{omega}{$\displaystyle\frac{\Omega}{\sqrt{G\rho_c}}$}
\psfrag{low}{$\Bav=1.44\times 10^{16}$ G}
\psfrag{med}{$\Bav=2.87\times 10^{16}$ G}
\psfrag{high}{$\Bav=5.71\times 10^{16}$ G}
\includegraphics[width=\linewidth]{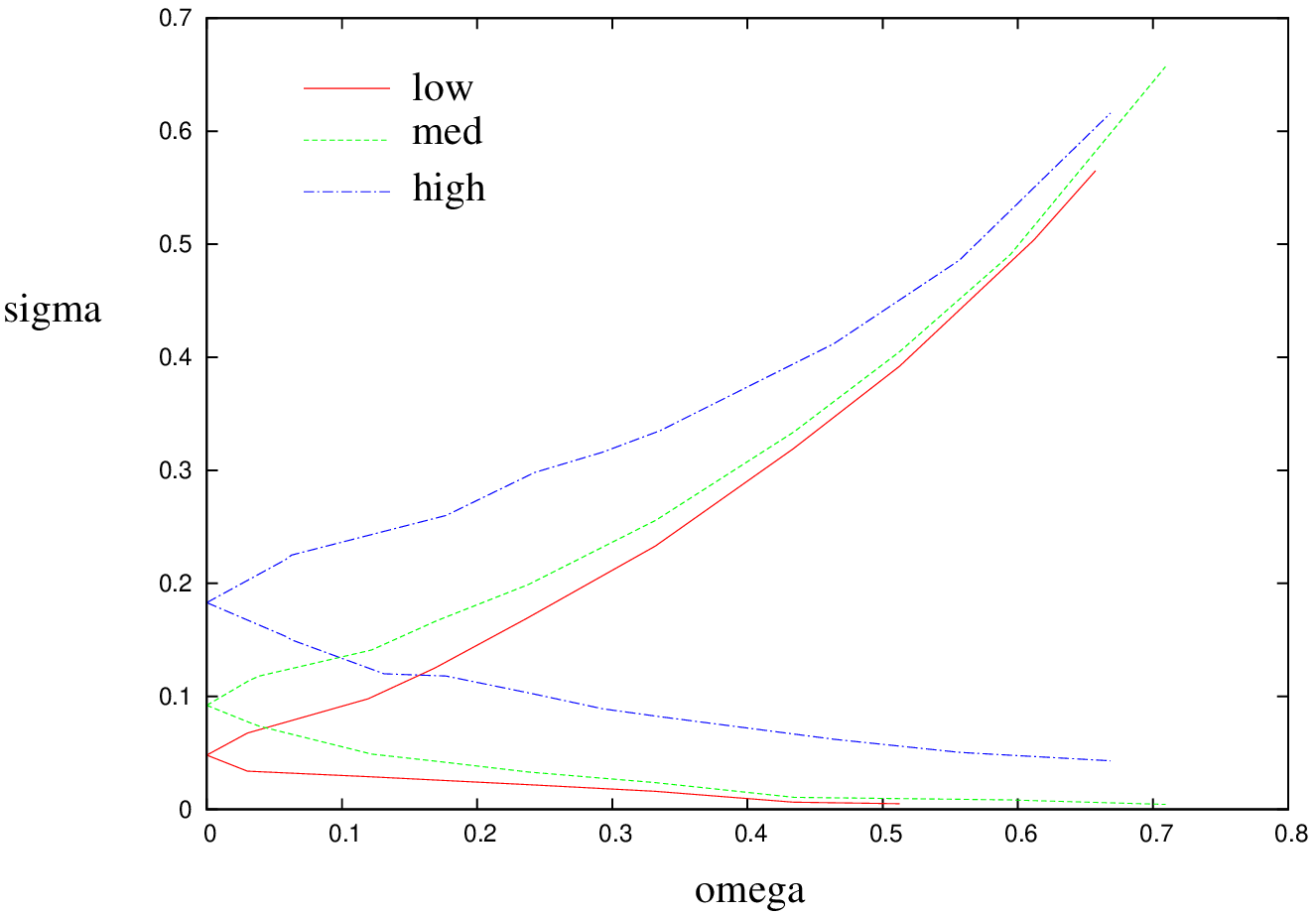}
\end{minipage}
\caption{\label{alf-to-r}
         Illustrating the hybrid inertial-magnetic nature of
         modes in a rotating magnetised star. When $\Omega=0$
         there is a pure $l=m=2$ $a$-mode, which is split into co- and
         counter-rotating modes by the effect of rotation. The
         counter-rotating mode frequency approaches the nonmagnetic
         $r$-mode frequency as $\Omega$ increases, while the
         corotating branch tends to zero frequency. The left-hand plot
         compares the ${}^2_2a$ mode with the $r$-mode, whilst
         the right-hand plot shows that the nature of the hybrid mode
         depends on the ratio $M/T$; when $\Bav$ is larger, the
         ${}^2_2a$-mode frequency approaches the $r$-mode frequency more
         slowly. Modes are tracked up to $\Omega\approx 0.7$ in
         dimensionless units, which is over 95\% of the break-up
         velocity. The irregular parts of the curves may correspond to
         avoided crossings with other magneto-inertial modes.}
\end{center}
\end{figure}

We begin by tracking the axial
${}^2_2a$-mode with increasing rotation, finding that as expected it
undergoes rotational splitting (figure \ref{alf-to-r}). The
lower-frequency branch of this 
$a$-mode appears to tend to zero with increasing $\Omega$ (or
equivalently, as $M/T\to 0$). The higher-frequency branch of
the $a$-mode tends to the $r$-mode frequency as $M/T\to 0$. We confirm
that the magnetic/inertial character of these hybrid modes depends on
$M/T$ by tracking the ${}^2_2a$-mode for three different field
strengths, finding that when $\Bav$ is higher the hybrid-mode
frequency approaches the $r$-mode frequency more slowly. The
higher-frequency branch of the ${}^2_2a$ mode is counter-rotating and
joins up with the $r$-mode, whilst the lower-frequency ${}^2_2a$ mode
corotates with the star.

\begin{figure}
\begin{center}
\begin{minipage}[c]{\linewidth}
\psfrag{sigma}{$\displaystyle\frac{\sigma}{\sqrt{G\rho_c}}$}
\psfrag{omega}{$\displaystyle\frac{\Omega}{\sqrt{G\rho_c}}$}
\psfrag{a1}{${}^4_2\!\!\:a_2$}
\psfrag{a2}{${}^4_2\!\!\:a_1$}
\psfrag{a3}{${}^4_2\!\!\:a_3$}
\psfrag{a4}{${}^3_2a_1$}
\psfrag{a5}{${}^3_2a_2$}
\psfrag{h1}{${}^4_2i_2$}
\psfrag{h2}{${}^4_2i_1$}
\psfrag{h3}{${}^4_2i_3$}
\psfrag{h4}{${}^3_2i_1$}
\psfrag{h5}{${}^3_2i_2$}
\includegraphics[width=\linewidth]{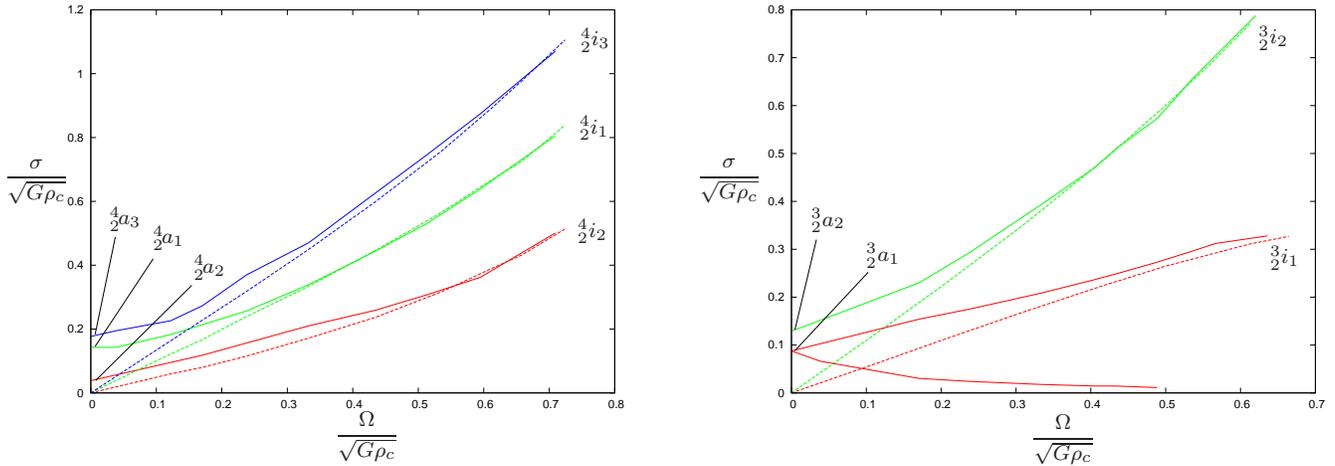}
\end{minipage}
\caption{\label{ah_modes}
         The $m\!=\!2,l_0\!=\!3,4$ hybrid magneto-inertial modes. Dashed
         lines represent the pure inertial ($\Bav=0$) modes, whilst solid
         lines show magneto-inertial modes, which reduce to pure
         Alfv\'en modes in the $\Omega\to 0$ limit. The left-hand plot
         shows the $l_0\!=\!4$ (axial) hybrid modes, whilst the
         right-hand plot shows $l_0\!=\!3$ (polar) modes. In each case
         the upper-frequency branch of a hybrid mode is seen to meet a
         corresponding $i$-mode as $M/T\to 0$. For the ${}^3_2a_1$ mode,
         we were also able to track the lower-frequency branch, which
         appears to reduce to a zero-frequency mode in the $M/T\to 0$ limit.}
\end{center}
\end{figure}

Having established that the pure ${}^2_2a$ mode and the pure $r$-mode
are replaced by a hybrid magneto-inertial mode when both magnetic and
rotational restoring forces are present, one would expect to find
similar hybrid modes corresponding to other Alfv\'en/inertial modes;
we confirm this expectation in figure \ref{ah_modes}. As before,
rotation splits each $a$-mode into co- and counter-rotating
branches. We are able to track the upper-frequency branches of both
polar ${}^3_2a$-modes to their inertial counterparts, and all three
${}^4_2a$-modes to known inertial modes in the $M/T\to 0$ limit. In
addition, we are able to track the lower-frequency branch of the
${}^3_2a_1$ mode to high rotation rates; it appears to become a
zero-frequency mode in the $M/T\to 0$ limit, as for the lower
${}^2_2a$-mode.

\begin{figure}
\begin{center}
\begin{minipage}[c]{\linewidth}
\psfrag{sigma}{$\displaystyle\frac{\sigma}{\sqrt{G\rho_c}}$}
\psfrag{Omega}{$\displaystyle\frac{\Omega}{\sqrt{G\rho_c}}$}
\psfrag{2a}{${}^2_2a$}
\psfrag{2r}{${}^2_2r$}
\psfrag{stable}{stable}
\psfrag{unstable}{unstable}
\psfrag{f_CFS}{$f_{\textrm{CFS}}(B)$ [Hz]}
\psfrag{Bav}{$\Bav$ [G]}
\includegraphics[width=\linewidth]{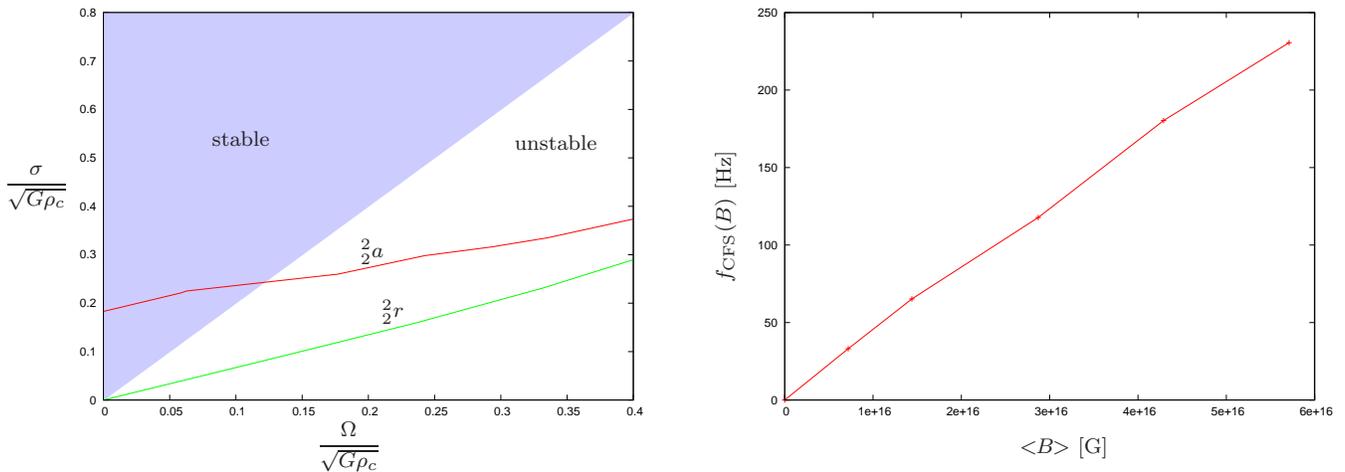}
\end{minipage}
\caption{\label{cfs_stab}
         In a slowly-rotating magnetised star, the $r$-mode is
         replaced by the axial $l=m$ $a$-mode. From the left-hand plot
         we see that this mode is not subject to the CFS instability
         if $\Omega$ is sufficiently small, but at some higher
         rotational frequency $f_{\textrm{CFS}}$ (a function of the
         field strength $B$) the $a$-mode crosses into the unstable
         regime. The right-hand plot shows the variation of
         $f_{\textrm{CFS}}$ with average field strength $\Bav$.}
\end{center}
\end{figure}

A number of modes are subject to instabilities driven by
gravitational radiation emission, but in general only become unstable
for sufficiently rapid rotation. However, the $r$-modes are unstable
even in very slowly rotating stars, in the absence of viscosity
\citep{nils_rmode}. We have already seen that magnetic
fields significantly alter the behaviour of the $r$-mode for slow
rotation, so we now consider the effect this has on their stability. For a
counter-rotating mode with frequency $\sigma$ (positive by convention)
in the rotating frame, the instability criterion is
\beq
\sigma(\sigma-m\Omega)<0.
\eeq
It follows immediately that radiative instabilities are suppressed
when $\sigma>m\Omega$. In the left-hand part of figure \ref{cfs_stab}
we plot this 
threshold frequency, together with the nonmagnetic $r$-mode and the
hybrid mode that replaces it in the magnetic case. It is clear that
whilst the unmagnetised $r$-mode is always in the unstable regime, its
magnetic equivalent (the hybrid of the $r$-mode and the axial $l=m$
$a$-mode) \emph{is} stable for sufficiently low rotation rates. The
maximum rotational frequency a star can have before its ${}^2_2r$ mode
goes unstable is plotted on the right-hand side, as a function of the
stellar magnetic field strength.

\begin{figure}
\begin{center}
\begin{minipage}[c]{0.5\linewidth}
\psfrag{omega}{$\displaystyle\frac{\Omega}{\sqrt{G\rho_c}}$}
\psfrag{tau}{\hspace{-0.5cm} $\displaystyle\frac{(\tau_{GR})_B}{(\tau_{GR})_0}$}
\includegraphics[width=\linewidth]{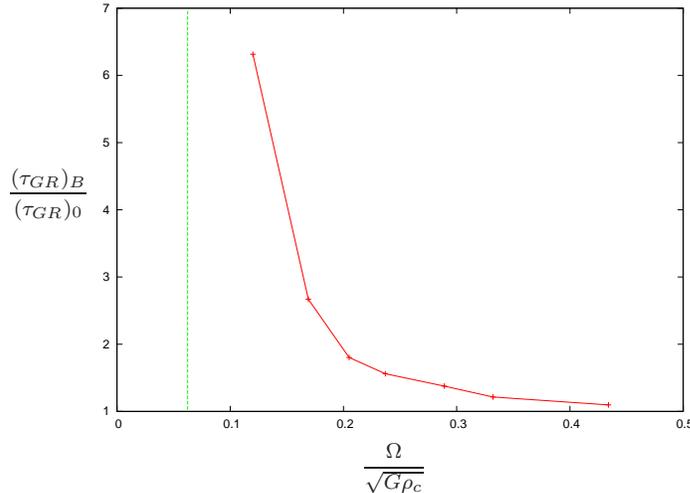}
\end{minipage}
\caption{\label{instab-growth}
         A magnetic field changes the growth time of the
         $r$-mode instability. Here we plot an approximation of the
         ratio of magnetised $(\tau_{GR})_B$ to unmagnetised
         $(\tau_{GR})_0$ growth timescales, against dimensionless
         rotation rate. The dashed vertical line shows where the ratio
         asymptotes (i.e. when the magnetised mode becomes stable). We see that 
         in all cases the instability growth is slower with magnetic
         effects, but the effect becomes insignificant for rapid
         rotation. The magnetic timescales used were for a star with a field
         of $2.87\times 10^{16}$ G.}
\end{center}
\end{figure}

Even when magnetic fields are not strong enough to suppress the
$r$-mode instability, they may slow down its growth. A full
calculation of this effect is beyond the scope of this paper, but we
may estimate it with some simplifying assumptions. The growth time
$\tau_{GR}$ of the $r$-mode instability due to gravitational radiation
is given by
\beq
\frac{1}{\tau_{GR}}=-\frac{1}{2E}\td{E}{t}
\eeq
where $E$ is the energy of the mode in the rotating frame. From this
one can show that the growth time $\tau_{GR}$ scales with the
rotating-frame mode frequency $\sigma$ in the following manner for an
$l=m$ $r$-mode:
\beq
\frac{1}{\tau_{GR}} \sim \sigma(\sigma-l\Omega)^{2l+1}
\eeq
--- see \citet{nils_rmode} for details. Note that for $l=m=2$, the
growth time scales with the sixth power of $\sigma$.

We wish to estimate how the growth time for the ${}^2_2r$-mode changes
when magnetic effects are included. Since $\tau_{GW}$ contains a
factor of $\sigma^6$, we will assume that this term has the most
significant variation when a magnetic field is added. Other terms in
the expression of $E$ and its derivative will be approximated as
constant. Using the indices $0$ and $B$ to denote nonmagnetic and
magnetic quantities (respectively), we then see that
\beq
\frac{(\tau_{GR})_B}{(\tau_{GR})_0}
   \approx  \frac{\sigma_0(\sigma_0-2\Omega)^5}{\sigma_B(\sigma_B-2\Omega)^5}.
\eeq
In figure \ref{instab-growth} we plot this dimensionless quantity as a
function of the rotation rate, finding that a toroidal magnetic field
does indeed slow down the instability's growth. The importance of the
effect depends on the rotation rate: at twice the threshold frequency
for stability of the magnetised $r$-mode (i.e. when the mode is
unstable), its growth time is still a factor of $\sim 6$ longer than
in the nonmagnetic $r$-mode case; however, for very rapid rotation the
difference in growth times is negligible.

\section{Discussion and conclusions}

In this paper we have investigated oscillation modes of
neutron stars with rotation and magnetic fields, specialising to the
case of purely toroidal background fields; we intend to study purely
poloidal and mixed-field geometries in future publications. Our
numerical approach allows us to study oscillations of rapidly rotating
and highly magnetised stars in a self-consistent manner. We first
generate a stationary star in equilibrium to use as the background
configuration; this star may have axisymmetric distortions due to 
magnetic effects and rotation. We then time-evolve linear perturbations
on this background star in order to study its modes of oscillation.

When a magnetic field is added to a star, the most obvious change to
its oscillation spectrum is the presence of Alfv\'en ($a$-) modes, a
class of stellar oscillation restored by the Lorentz force. These
modes are purely magnetic in nature only for a nonrotating background
star.

In a rotating magnetised star, we find that the pure $a$-modes of a
nonrotating star (or equivalently, the purely inertial $i$-modes of an
unmagnetised star) are replaced by hybrid magneto-inertial modes,
whose character is governed by the ratio of the magnetic $M$ and
kinetic $T$ energies, as discussed by \citet{morsink}. Tracking a star
at fixed magnetic field from 
$\Omega=0$ through increasing rotation rate, we see a rotational
splitting of the $a$-modes into co- and counter-rotating modes. The
higher-frequency branches of these modes approach known $i$-mode
frequencies. In general the lower-frequency branches are harder to track,
owing to the dense nature of the oscillation spectrum, but when we are
able to identify them we find that they appear to become
zero-frequency modes in the $M/T\to 0$ limit.

The presence of these hybrid modes has parallels with other work. The
evolutions of \citet{passa_strat} and \citet{gaertig} found that when
tracking $g$-modes 
(i.e. modes restored by composition gradients within the star) for
increasingly rapid rotation, their frequencies approached known
$i$-mode frequencies. One key difference between stratified and
magnetised stars, however, is the behaviour of the $r$-mode in each
case. Being purely axial in the slow-rotation limit, the $r$-modes are
unaffected by composition gradients, whereas we have found that the
presence of a magnetic field means that in the slow-rotation limit
they become the axial $l=m$ $a$-mode.

Our work seems to be
consistent with the analysis of \citet{glam_ander}, who found that
magnetic fields could act to suppress instabilities driven by
gravitational radiation (the CFS instability); and in particular, that
purely poloidal or 
purely toroidal fields should always play a stabilising role in this
case. Using $\sigma$ to denote a mode frequency as 
measured in the rotating frame, it is known that modes satisfying
the condition $\sigma(\sigma-m\Omega)<0$ are susceptible to these
radiation-driven instabilities; in particular, this includes the
$r$-mode. In the presence of a magnetic field we find that the
$r$-mode is replaced by the $l=m$ axial $a$-mode; for sufficiently
slow rotation we have $\sigma_a>m\Omega$ and hence the mode is
CFS-stable. In the regime where the star is unstable, we use a simple
estimate to suggest that the instability's growth will be slower in
the presence of a magnetic field.

In addition to the hybrid magneto-inertial modes, there are also
magnetic corrections to the $f$-mode frequency. These corrections are
very modest ($\sim$1\%) even up to field strengths of the order
$10^{17}$ gauss. In addition, as for the magneto-inertial modes, the
magnetic correction becomes less significant still as $M/T\to
0$. However, we note that the magnetic correction seems to increase
between $m=2$ and $m=6$, so although our approach limits us to low $m$
one might expect more appreciable corrections to high-$m$ $f$- and
$p$-modes.

Although it would be premature to make a quantitative comparison between
our results and observed magnetar QPOs, we note that there are certain
similarities in the oscillation spectra. The QPOs observed from SGR
1806-20 include 26 and 30 Hz modes; these cannot be explained as
overtones of crustal shear modes because the spacing is too small
(they would need to be integer multiples for this). By contrast, it is
easy to interpret these frequencies as global modes of a fluid star
(i.e. the magnetar's interior), since we see modes at far smaller
separation than integer multiples. For example, using our fitted
relation \eqref{linear-sigB} we see that the frequency ratio of the axial
${}^2_2a$ and polar ${}^3_2a_1$ modes is $0.030/0.033\approx 0.91$ ---
comparable with the observed ratio of $26/30\approx 0.87$.

This paper adds to the picture of magnetic stellar oscillations built
up by a number of other recent numerical studies. The work of
\citet{sotani08} and \citet{cerda} 
investigated axial magnetar oscillations, modelling the star's
magnetic field as dipolar (and hence purely poloidal). They found two
localised families of QPOs, which they related to observed magnetar
QPOs. \citet{colaiuda} worked on a similar problem, but in the more
general case of a mixed poloidal-toroidal background field. Their work
complements other studies, but they are also able to identify a third
family of QPOs in their model star. Finally, \citet{sotani09} find a
set of polar oscillations of dipolar fields, agreeing with the work of
\citet{uminlee} that a magnetar should have both axial and polar
oscillations.

Many of these recent studies have analysed their results in the light
of the suggestion that magnetic oscillations of a perfectly-conducting
star form a continuum, rather than discrete modes. This was proposed
by \citet{levincont}, revisiting earlier work by
\citet{goos_cont}. Various numerical studies
\citep{sotani08,cerda,colaiuda} have found results consistent with
this proposal, in the case of axial oscillations of a dipole
field. However, \citet{sotani09} suggest that \emph{polar}
oscillations of a dipolar-field star are discrete.

Since our background field is purely toroidal, we cannot make
quantitative comparisons with work discussed in the last two
paragraphs, since those studies assumed dipolar fields (or mixed
poloidal-toroidal fields in the case of \citet{colaiuda}). However, we
do find broad similarities --- in particular, our $a$-mode frequencies
are of the order 100 Hz (for a field of $\sim 10^{16}$ gauss), as
found from other magnetic 
evolutions. With a toroidal field there is no reason to expect a
continuum of modes, since this feature has only be established for
fields with a poloidal component. Indeed, all our results have shown
discrete mode frequencies, with no dependence on position within the
star, up to uncertainties due to resolution and the finite duration of
our simulations (in practice, errors of $\sim 1\%$). Our polar
$a$-modes thus share this property with those of 
\citet{sotani08}, but our axial $a$-modes are discrete too.

Purely toroidal fields and purely poloidal fields suffer from generic
localised instabilities, so in the absence of damping mechanisms are
not viable candidates for long-lived stellar magnetic fields
\citep{wright,taylertor}. Despite this, we have been able to perform
stable evolutions of perturbations about a purely toroidal background
for this work. There may be a number of reasons why these
analytically-established instabilities have not affected our numerical
work. Since we only consider first-order perturbations, higher-order
effects are avoided; at the linear level, the greatest instabilities
are those for $m=0$ and $m=1$, whilst we have only considered $m\geq
2$. Finally, we have included artificial viscosity and resistivity to
damp numerically-generated instabilities, and it is possible that
these have prevented the growth of physical instabilities too.

One way in which pure-poloidal/toroidal fields may be stabilised is
through rotation \citep{geppert,braithtor,kiuchi_evol}, although this
effect will be small in the case of the magnetars, whose rotational
periods are very long. Relatively small poloidal components may
stabilise dominantly toroidal fields \citep{braithtorpol}, but it is
difficult to draw general conclusions on the relative strengths of the
two components, since other work \citep{landerjones,ciolfi} has found
that apparently general constructions of magnetic stars in equilibrium
(in both Newtonian and relativistic contexts) result in mixed
fields which are dominantly \emph{poloidal}.

Given the many uncertainties regarding the nature of stellar magnetic
fields, we believe that it is reasonable to study oscillations of
purely toroidal fields, even though these may suffer certain
instabilities, as we have discussed. Furthermore, a star whose field
is dominantly toroidal could be expected to have an oscillation
spectrum with similar features to those discussed in this paper.

The observations of QPOs in the tails of giant flares from magnetars
may be providing a rare probe of the interior fields of these stars. In
addition, the longer-term prospect of studying gravitational-wave
signals from neutron stars using detectors like Advanced LIGO and
VIRGO should also help understand the physics of these stars. With a
clearer picture of the nature of stellar oscillations in a strong
magnetic field, we should be better equipped to interpret magnetar
QPOs and GW signals of neutron stars. Our formalism allows us to
build up a more sophisticated picture of the oscillation spectrum of
magnetised stars by a gradual inclusion of extra effects. These could
include other field geometries, stratified stars, a superfluid
interior, and so on; we intend to explore these in future work.

\section*{Acknowledgments}

We thank Nils Andersson for useful discussions and advice. This work
was supported by STFC through grant number PP/E001025/1 and by
CompStar, a Research Networking Programme of the European Science
Foundation.

\label{lastpage}


\begin{thebibliography}{99}
\bibitem[\protect\citeauthoryear{Andersson \& Kokkotas}{2001}]{nils_rmode}
  Andersson N., Kokkotas K.D., 2001, International Journal of Modern
  Physics D, 10, 381
\bibitem[\protect\citeauthoryear{Brackbill \& Barnes}{1980}]{brackbarnes}
  Brackbill J.U., Barnes D.C., 1980, J. Comp. Phys., 35, 426
\bibitem[\protect\citeauthoryear{Braithwaite}{2006}]{braithtor}
  Braithwaite J., 2006, A\&A, 453, 687
\bibitem[\protect\citeauthoryear{Braithwaite}{2009}]{braithtorpol}
  Braithwaite J., 2009, MNRAS, 397, 763
\bibitem[\protect\citeauthoryear{Cerd\'a-Dur\'an et al.}{2009}]{cerda}
  Cerd\'a-Dur\'an P., Stergioulas N., Font J.A., 2009, MNRAS, 397, 1607
\bibitem[\protect\citeauthoryear{Chandrasekhar \& Limber}{1954}]{chandlimb}
  Chandrasekhar S., Limber D.N., 1954, ApJ, 119, 10
\bibitem[\protect\citeauthoryear{Ciolfi et al.}{2009}]{ciolfi}
  Ciolfi R., Ferrari V., Gualtieri L., Pons J.A., 2009, MNRAS, 397,
  913
\bibitem[\protect\citeauthoryear{Colaiuda et al.}{2009}]{colaiuda}
  Colaiuda A., Beyer H., Kokkotas K.D., 2009, MNRAS, 396,
  1441
\bibitem[\protect\citeauthoryear{Cox}{1980}]{cox_book}
  Cox J.P., 1980, Theory of Stellar Pulsation, Princeton University Press
\bibitem[\protect\citeauthoryear{Dedner et al.}{2002}]{dedner}
  Dedner A., Kemm F., Kr\"oner D., Munz C.-D., Schnitzer T., Wesenberg
  M., 2002, J. Comp. Phys., 175, 645
\bibitem[\protect\citeauthoryear{Dziembowski \& Goode}{1996}]{dziem}
  Dziembowski W.A., Goode P.R., 1996, ApJ, 458, 338
\bibitem[\protect\citeauthoryear{Friedman \& Schutz}{1978}]{friedmanschutz}
  Friedman J.L., Schutz B.F., 1978, ApJ, 221, 937
\bibitem[\protect\citeauthoryear{Gaertig \& Kokkotas}{2009}]{gaertig}
  Gaertig E., Kokkotas K.D., 2009, PRD, 80, 064026
\bibitem[\protect\citeauthoryear{Geppert \& Rheinhardt}{2006}]{geppert}
  Geppert U., Rheinhardt M., 2006, A\&A, 456, 639
\bibitem[\protect\citeauthoryear{Glampedakis \& Andersson}{2007}]{glam_ander}
  Glampedakis K., Andersson N., 2007, MNRAS, 377, 630
\bibitem[\protect\citeauthoryear{Goossens et al.}{1985}]{goos_cont}
  Goossens M., Poedts S., Hermans D., 1985, Solar Phys., 102, 51
\bibitem[\protect\citeauthoryear{Israel et al.}{2005}]{israel}
  Israel G.L. et al., 2005, ApJ, 628, L53
\bibitem[\protect\citeauthoryear{Jones et al.}{2002}]{jones_evol}
  Jones D.I., Andersson N., Stergioulas N., 2002, MNRAS, 334, 933
\bibitem[\protect\citeauthoryear{Kiuchi et al.}{2008}]{kiuchi_evol}
  Kiuchi K., Shibata M., Yoshida S., 2008, PRD, 78, 024029
\bibitem[\protect\citeauthoryear{Lander \& Jones}{2009}]{landerjones}
  Lander S.K., Jones D.I., 2009, MNRAS, 395, 2162
\bibitem[\protect\citeauthoryear{Lee}{2007}]{uminlee}
  Lee U., 2007, MNRAS, 374, 1015
\bibitem[\protect\citeauthoryear{Ledoux \& Simon}{1957}]{ledoux}
  Ledoux P., Simon R., 1957, AnAp, 20, 185
\bibitem[\protect\citeauthoryear{Levin}{2007}]{levincont}
  Levin Y., 2007, MNRAS, 377, 159
\bibitem[\protect\citeauthoryear{Lockitch \& Friedman}{1999}]{lock_fried}
  Lockitch K.H., Friedman J.L., 1999, ApJ, 521, 764
\bibitem[\protect\citeauthoryear{Morsink \& Rezania}{2002}]{morsink}
  Morsink S.M., Rezania V., 2002, ApJ, 574, 908
\bibitem[\protect\citeauthoryear{Passamonti et al.}{2009}]{passa_strat}
  Passamonti A., Haskell B., Andersson N., Jones D.I., Hawke I.,
  2009, MNRAS, 394, 730
\bibitem[\protect\citeauthoryear{Price \& Monaghan}{2005}]{pricemona}
  Price D.J., Monaghan J.J., 2005, MNRAS, 364, 384
\bibitem[\protect\citeauthoryear{Rincon \& Rieutord}{2003}]{rinc_rieu}
  Rincon F., Rieutord M., 2003, A\&A, 398, 663
\bibitem[\protect\citeauthoryear{Sotani et al.}{2008}]{sotani08}
  Sotani H., Kokkotas K.D., Stergioulas N., 2008, MNRAS, 385, L5
\bibitem[\protect\citeauthoryear{Sotani \& Kokkotas}{2009}]{sotani09}
  Sotani H., Kokkotas K.D., 2009, MNRAS, 395, 1163
\bibitem[\protect\citeauthoryear{Stergioulas et al.}{2004}]{ster_eigenfn}
  Stergioulas N., Apostolatos T.A., Font J.A., 2004, MNRAS, 352, 1089
\bibitem[\protect\citeauthoryear{Tayler}{1973}]{taylertor}
  Tayler R.J., 1973, MNRAS, 161, 365
\bibitem[\protect\citeauthoryear{Unno et al.}{1989}]{unno}
  Unno W., Osaki Y., Ando H., Saio H., Shibahashi H., 1989,
  Nonradial Oscillations of Stars, University of Tokyo Press
\bibitem[\protect\citeauthoryear{Watts \& Strohmayer}{2007}]{watts_stroh}
  Watts A.L., Strohmayer T.E., 2007, Adv. Space Res., 40, 1446
\bibitem[\protect\citeauthoryear{Wickramasinghe \& Ferrario}{2000}]{wickrama}
  Wickramasinghe D.T., Ferrario L., 2000, PASP, 112, 873
\bibitem[\protect\citeauthoryear{Wright}{1973}]{wright}
  Wright G.A.E., 1973, MNRAS, 162, 339
\end{thebibliography}
\end{document}